\documentstyle[osa]{revtex}
\begin{document}
%\preprint{Version IOTA, \today}
%\draft

\newcommand{\beq}{\begin{equation}}
\newcommand{\eeq}{\end{equation}}
\newcommand{\ben}{\begin{eqnarray}}
\newcommand{\een}{\end{eqnarray}}
\newcommand{\bea}{\begin{array}}
\newcommand{\eea}{\end{array}}
\newcommand{\om}{(\omega )}
\newcommand{\bef}{\begin{figure}}
\newcommand{\eef}{\end{figure}}
\newcommand{\leg}[1]{\caption{\protect\rm{\protect\footnotesize{#1}}}} 

\newcommand{\ew}[1]{\langle{#1}\rangle}
\newcommand{\be}[1]{\mid\!{#1}\!\mid}
\newcommand{\no}{\nonumber}
\newcommand{\etal}{{\em et~al }}
\newcommand{\geff}{g_{\mbox{\it{\scriptsize{eff}}}}} 
\newcommand{\da}[1]{{#1}^\dagger}
\newcommand{\cf}{{\it cf.\/}\ }
\newcommand{\ie}{{\it i.e.\/}\ }

\title{  \center{Photon antibunching in the fluorescence 
of individual colored centers in diamond}}

\author{Rosa Brouri, Alexios Beveratos, 
Jean-Philippe Poizat, and Philippe Grangier}
\address{Laboratoire Charles Fabry de l'Institut  d'Optique, UMR 8501 du CNRS, \\
B.P. 147, 
 F91403 Orsay Cedex - France\\
e-mail : jean-philippe.poizat@iota.u-psud.fr}
\maketitle
%
%\centerline{\today}
%
\begin{abstract}
We have observed photon antibunching in the fluorescence light
emitted from a single N-V center in diamond at room temperature.
The possibility of generating triggerable single photons
with such a solid state system is discussed.

\vspace{0.5cm}

OCIS numbers: 270.5290 - 180.1790 - 030.5260
\vspace{0.5cm}

\end{abstract}

The security of quantum cryptography is based on the fact that
each bit of information is coded on a single quantum object, 
namely a single photon. The fundamental impossibility of duplicating the complete 
quantum state of a single particule prevents any potential eavesdropper from 
intercepting the message without the receiver noticing \cite{TRG}. In this context,
the realization of a efficient and integrable light source delivering a train of 
pulses containing one and only
one photon appears as an extremely challenging goal \cite{L}.
For reaching it, many issues should be addressed, which
range from achieving the full control of the quantum properties of the source
to an easy handling and integrability in a practical quantum cryptography set-up.

Several pioneering experiments have been realized for obtaining
single photon sources \cite{HM,GRA,MGM,KBKY,BLTO}.
Refs \cite{HM,GRA} are based on the  spontaneous
emission of twin photons. One of the  photon is used to open 
a time gate during which the twin photon can be observed. In these types of set-ups,
 the emission time of the photon is random. 
In quantum cryptography based communication systems  
it is desirable to deal with single  photons synchronized on an external clock, namely
triggerable single photons \cite{MGM,KBKY,BLTO}.
Experiment by Kim et al \cite{KBKY} uses the regulation of the 
electron flux at the level of a single electron 
in a  light emitting mesoscopic p-n junction
to realize a single photon turnstile device.
Experiments by de Martini et al \cite{MGM} and Brunel et al \cite{BLTO}
are based on the collection of fluorescence light emitted
by a single dye molecule. 
The experiments reported in \cite{KBKY,BLTO} require cryogenic 
apparatus (liquid He temperature).
The experiment reported in \cite{MGM} may run at room temperature, 
but then the molecules are rapidly destroyed by photobleaching
and must be renewed, e.g. by having them in a liquid solvant \cite{KJRT}. 
Another general limitation in these
experiments is the very small collection efficiency ($\eta < 10^{-3}$),
which means that less than $0.1 \%$
of the emitted photons are actually detected.

Generally speaking, the physical principle of single-molecule sources is that 
a single emitting dipole will emit only one photon at a time.
Using an adequate pulsed excitation scheme, it is then possible to produce
only one photon per pulse\cite{MGM,LK}.
In this letter, we report on the
use of single colored centers in diamond 
as single quantum emitters \cite{GDTFWB}. 
By investigating the quantum statistics of the fluorescence light
we have observed photon antibunching, which
is a clear signature of the unicity of the emitting dipole.
Photon antibunching has already been observed in many experiments,
involving e.g. a single atom \cite{KDM}, a single
trapped ion \cite{DW}, and a single molecule \cite{BMOT}.
With respect to these experiments, 
we point out that our set-up is particularly simple, 
since it involves bulk diamond at room temperature, 
and a non-resonant excitation from an argon-ion laser at 514 nm,
with a typical power of 10 mW. 
The present collection efficiency is still low (see discussion below),
but straightforward improvements are possible.
We consider thus that the present result is a
first step towards the realization of a simple and efficient
solid state source for single photons.

The colored center that has been used is the Nitrogen-Vacancy (N-V) defect center
in diamond, with a zero phonon line at a wavelength of $637$nm.
The defect consists in a substitutionnal nitrogen and a vacancy in a adjacent 
site. A simplified level structure is a four-level scheme 
with fast non radiative decays within the two upper states
and within the  two lower states. The excited state
lifetime  is $\tau =11.6$ ns \cite{CTJ}. 
We have used $0.1 \times 1.5 \times 1.5 $mm$^3$ single [110]
crystals of synthetic Ib
diamond from Drukker International.
Nitrogen is present in the crystal as an impurity.
Vacancies are created by irradiation with $2$ MeV electrons at a dose
of $3\times 10^{12} e^{-}/$cm$^2$. 
The irradiation dose is chosen so that the
density of vacancies is of the order of 1 $\mu$m$^{-3}$.
After irradiation the  crystals are annealed in vacuum
at $850^o$C during $2$ hours to form the N-V centers.
A remarkable property of these centers is that we could not
observe any photobleaching : the fluorescence level remains unchanged
after several hours of continuous
laser irradiation of a single center in the saturation regime.
On the other hand, a limitation of the system is the existence of shelving 
in a metastable  singlet state. This leads to  the observation of photon bunching
for time scale longer than the lifetime of the center  \cite{KJRT,BMOT,BFTO},
and to a decrease of the fluorescence count rate, owing to the time 
spent in this long lived state. Various
techniques for laser-assisted deshelving have been proposed in the 
literature \cite{DFTJKNW}. Such mechanisms should be eventually
considered, in relation with a pulsed excitation scheme
that will be required for true single-photon generation.

The experimental set-up is depicted in Fig. \ref{fig-exp}.
It is based upon a home-made scanning confocal microscope. 
The green line ($\lambda = 514$ nm) of a Ar$^+$ cw laser
is focussed on the sample by a high numerical aperture (1.3) 
immersion objective (Nachet 004279). A PZT-mounted
mirror located just before the objective allows for a x-y
scan of the sample, and fine z-scan is obtained using another PZT.
The fluorescence (wavelength between $637$ and $800$ nm)
is collected by the same objective
and separated from the excitation laser by a dichroic mirror. 
High rejection rate ($10^{15}$) long pass filters are used to removed any 
green or blue light. The confocal operation is
achieved by imaging the sample onto a $50 \mu $m pinhole 
with a magnification of $100$.
In order to investigate the fluorescence intensity correlation
we use an Hanbury-Brown and Twiss setup with two 
avalanche photodiodes (EG$\&$G, model SPCM-AQR 13) and a $50/50$
beamsplitter. IR filters and pinholes are set in front of each detector
to prevent optical cross-talk  between them
\cite{crosstalk}.
The time delay between pulses from the two photodiodes
is converted by a time to amplitude converter (TAC) into a voltage amplitude,
that is digitalized by the data acquisition card of a computer.
A delay of $40$ ns in one TAC input allows the measurement of
the intensity correlation for negative time. 

Fig. \ref{scan} shows a scan of the sample. 
Individual bright spots corresponding to N-V centers
appear clearly on the scan, with an apparent lateral resolution of $500$ nm. 
However, we note that 
the scan has been obtained at a depth of $10$ $\mu$m below the diamond surface,
where the spherical aberration caused by the diamond-oil interface
is not negligeable (the refractive index of diamond is $n=2.4$). 
An evaluation done using a lens-design software
indicates that only $20 \%$ of the emitted light is actually in the 
main peak, while about $80 \%$ is spread out in a halo which is about 
$3$ $\mu$m in diameter. 
Similar aberrations appear on the focused laser beam. 
For an input power of 15 mW, 
the typical signal from an individual center is $S_{det}= 2000$ s$^{-1}$
on each photodiode. 
The evaluated overall detection efficiency  is $\eta_{tot} = 0.0014$.
It can be split in
$\eta_{geom} = 0.08$ (geometrical collection angle),
$\eta_{ab} = 0.2$ (aberrations due to the diamond-oil interface),
$\eta_{opt} = 0.25$ (optical transmission),
$\eta_{BS} = 0.5$ (beamplitter), and
$\eta_{det} = 0.7$ (detector quantum efficiency).
The inferred emission 
rate from the center is thus of  $S_{det}/\eta_{tot}=S_{em}= 
1.4 \times 10^6$ s$^{-1}$.
The difference between $S_{em}$ and the fully saturated value
$S_{rad}= 9 \times 10^7$ s$^{-1}$ is attributed to the joint effects of
shelving, and of a non-saturated excitation. 
It can also be seen from Fig. \ref{scan} that the peaks 
appear above a background level with a typical value $B=4000$ s$^{-1}$. 
When increasing the incident laser power (in the range 2-20 mW), 
a saturation behaviour
appears on the signal fluorescence, while the background level increases
linearly. The intensity correlation
from the signal and from the background also behave differently, as it
will be discussed in more detail below.

The raw coincidences $c(t)$ (right axis) 
and the correlation function $g^{(2)}(t)$ (left axis) are 
represented in Fig. \ref{antibunching}. 
There $t$ is the delay between two joint photodetection events.
In order to have a good statistics on the correlations,
the typical data acquisition time $T$ is a few hours.
A slow (8 s response time) x-y-z computerized servo-lock was used to maintain 
the fluorescence on the maximum of the N-V center under study.
Short-term and long-term drifts of the laser intensity are less than $10 \% $. 

For a Poissonian light source the coincidence rate 
(in s$^{-1}$) in a time bin of width $w$ is
$N_1 N_2 w$, where $N_{1,2}$ are the count rates on each detector.
The raw coincidence number $c(t)$ is normalized to the one of a Poissonian source
 according to
\beq
C_N (t) = c(t) /(N_1 N_2 w T) \; .
\eeq
By stopping the scan between several centers
(typically 1 $\mu$m away from a center of interest), 
we have checked experimentally 
that the normalized coincidence rate from the background light is 
flat and equal to unity. We thus make the reasonable assumption that
the photons emitted from the background are uncorrelated with the
photons emitted from the center at the same location.
This allows us to correct for the random
coincidences caused by the background light and obtain the $g^{(2)}(t)$ 
correlation function of the N-V center : 
\beq
g^{(2)} (t)= (C_N(t) - (1 - \rho ^2))/\rho^2
\eeq
where $\rho = S/(S+B)$ is related to the signal to background ratio,
which is measured independantly in each experimental run.
After this substraction procedure,
the $g^{(2)}(t)$ correlation function shown in Fig. \ref{antibunching}(b)
goes to zero for $t=0$. This is consistent with our claiming 
that there is only a single N-V center in the
fluorescence peak selected in Fig. \ref{scan}. 
In the case of the presence of $n$ spatially unresolved centers,
the value of the zero-time antibunching is $1-1/n$.
Experimental runs with $g^{(2)}(0) > 0$ were also commonly observed, and
are  typical of the brightest spots seen on Fig. \ref{scan}. 
In Fig. \ref{antibunching}(b), the non-unity value of 
$g^{(2)}(t)$ for $t > \tau $ is the bunching effect \cite{KJRT,BMOT,BFTO}
owing to the presence of a metastable singlet state \cite{DFTJKNW}.
We have observed experimentally 
that this bunching effect depends on the laser
power, indicating that the 514 nm excitation light may also contribute
to deshelving the metastable state.  
A quantitative study of these effects is under way.

In the present experiment the  background level ($\rho = 0.34 $) 
is too high for an efficient use in  quantum cryptographic systems. 
Spectrographic analysis of the 
fluorescence light reveals that significant improvement should be 
obtained by optimizing the spectral filters 
and dichroic mirrors, in order to 
eliminate stray fluorescence around 600 nm which does not arise from the NV 
center. Preliminary measurements show that
the fraction of useful signal $\rho$ can then reach values above $0.8$.
Other improvements can also be expected from a non-aberrating 
collection optics, from truely fluorescence-free immersion oil,
and from an optimized choice of the diamond sample. 

As a conclusion, we have observed photon antibunching from a very simple
set-up involving a diamond crystal at room temperature.
The total absence of photobleaching allowed us to lock
the laser beam on a single center during several hours. 
Colored centers
are included in a solid matrix, easy to handle, and they appear as
good candidates for realizing single photon sources
for quantum cryptography. 
Ultimate efficiency for such a source should be obtained by 
coupling the emitting dipole to a microcavity, in order to emit the light
in a single mode \cite{LK}. We point out however that 
a possible alternative for
achieving overall efficiencies in the range $10-20 \%$
may be the use of optimized wide-aperture collection optics.

We thank Hans Wilhelm for enlightening 
discussions at the initial steps of this experiment.
We also thank E. Br\'eelle and M. Vidal from  the ``Groupe de Physique des Solides''
at Paris 6 for the sample irradiation, and A. Machu for sample annealing.
This work is supported by the European IST/FET program
``Quantum Information Processing and Telecommunication'', 
project number 1999-10243 ``S4P".

%%%%%%%%%%%%%%%%%%%%%%%%%%%%%%%%%%%%%%%%%%%%%%%%%%%%%%%%%%%%%%%%%%%%

%%%%%%%%%%%%%%%%%%%%%%%%%%%%%%%%%%%%%%%%%%%%%%%%%%%%%%%%%%%%%%%%

\begin{figure}
\caption{Experimental set-up. The sample fluorescence is excited and 
collected using a confocal microscope set-up. The intensity correlations are
measured using two avalanche photodiodes, a time to amplitude converter and 
a multichannel analyzer.}
\label{fig-exp} 
\end{figure}

\begin{figure}
\caption{(a) Confocal microscopy raster scan 
($ 5 \; \times \;  5 \; \mu$m$^2$) of the sample
performed about $10  \; \mu$m$$ below the diamond surface.
The size of a pixel is $60$ nm. The integration time per pixel is $32$ ms.
The laser intensity impinging on the sample is $15$ mW.
(b) Line scan along the dotted line in (a). The data is shown
together with a gaussian fit,
which is used to evaluate the signal and background levels.
Here we obtain $\rho = S/(S+B) = 0.34$}
\label{scan} 
\end{figure}

\begin{figure}
\caption{Normalized correlation function $g^{(2)}(t)$ corrected for the
random coincidences from the background ($\rho = 0.34$, see text).
The data corresponds to the center circled in Fig. \protect\ref{scan}(a).
The actual number of coincidences is indicated on the right. 
The time bin is $w=1$ ns, the integration time is $T=11450$ s, and the single
count rates are $N_1=5780$ s$^{-1}$, $N_2=5990$ s$^{-1}$.
The full line is an exponential fit to the data,
using the model described in ref. \protect\cite{KJRT}.}
\label{antibunching} 
\end{figure}

\end{document}